\documentclass[useAMS]{mn2e}
\usepackage{lscape}
\usepackage{rotating}

\def\Msun{\hbox{M$_\odot$}}
\def\Ne{\hbox{$N_{\rm {e}}$}}
\def\Te{\hbox{$T_{\rm {e}}$}}
\def\kms{\hbox{km$\,$s$^{-1}$}}
\def\cm3{\hbox{cm$^{-3}$}}
\def\one{\,{\sc i}}             
\def\two{\,{\sc ii}}
\def\three{\,{\sc iii}}

\input psfig.sty
\input epsf.sty
\usepackage{graphicx}
\title[HST/STIS Optical Spectroscopy of SSCs in M82]
{HST/STIS Optical Spectroscopy of Five Super Star Clusters
in the Starburst Galaxy M82}

\author[L. J. Smith et al.] {L. J. Smith$^1$, M.S. Westmoquette$^1$,
J. S. Gallagher III$^2$, R.W. O'Connell$^3$, 
\newauthor D.J.
Rosario$^3$ and R. de Grijs$^4$\\
$^1$Department of Physics and
Astronomy, University College London, Gower Street, London, WC1E 6BT\\
$^2$Department of Astronomy, University of Wisconsin-Madison, 5534
Sterling, 475 North Charter St., Madison WI 53706, USA\\
$^3$Department of Astronomy, University of Virginia, 
P.O. Box 3818, Charlottesville, VA 22903, USA\\
$^4$Department of  Physics and Astronomy, The University of Sheffield,
Hicks Building, Hounsfield Rd., Sheffield, S3 7RH
}
\date{Accepted 2006 April 27. Received 2006 April 11; in original form 2006 February 11}
\pagerange{\pageref{firstpage}--\pageref{lastpage}}
\pubyear{2006}
\begin{document}
\maketitle
\label{firstpage}
\begin{abstract}
We present optical spectroscopy obtained with the Space Telescope Imaging Spectrograph (STIS) of five young massive star clusters in the starburst galaxy M82.  A detailed analysis is performed for one cluster `M82-A1' and its immediate environment in the starburst core. From HST archive images, we find that it is elliptical with an effective radius of $3.0\pm0.5$ pc and is surrounded by a compact ($r=4.5\pm0.5$ pc) H\two\ region. We determine the age and reddening of M82-A1 using synthetic spectra from population synthesis models by fitting both the continuum energy distribution and the depth of the Balmer jump. We find an age of $6.4\pm0.5$~Myr and a photometric mass estimate of $M=7$--$13 \times 10^{5}$ M$_{\odot}$. We associate its formation with the most recent starburst event 4--6 Myr ago. We find that the oxygen abundance of the H\two\ region surrounding M82-A1 is solar or slightly higher. The H\two\ region has a high pressure $P/k = 1$--2$ \times 10^{7}$~\cm3\,K. The diffuse gas in region A has a slightly lower pressure, which together with the broad H$\alpha$ emission line width, suggests that both the thermal and turbulent pressures in the M82 starburst core are unusually high.  We discuss how this environment has affected the evolution of the cluster wind for M82-A1. We find that the high pressure may have caused the pressure-driven bubble to stall. We also obtain spectroscopic ages for clusters B1-2 and B2-1 in the `fossil' starburst region and for the intermediate age clusters F and L. These are consistent with earlier studies and demonstrate that star formation activity, sufficiently intense to produce super star clusters, has been going on in M82 during the past Gyr, perhaps in discrete and localized episodes.
\end{abstract}
\section{Introduction}\label{intro}
Young, massive super star clusters (SSCs)\footnote {In this paper, we
define SSCs to include star clusters with half-mass radii of $\le
5$~pc; ages of $\le 500$~Myr; and total masses of $\ge 10^{5}$~\Msun}
are a prominent product of the intense star formation episodes that
occur in nearby starbursts, and presumably in merging galaxies at high
redshift. They are important objects to study because of the insights
they provide on: the formation and destruction of globular clusters;
star formation processes in extreme environments; and the triggering
and feeding of supergalactic winds.  The properties of young massive
star clusters are extensively reviewed in a recent proceedings
(Lamers, Smith \& Nota 2004).
 
M82 is the best nearby example of a giant starburst galaxy at a
distance of 3.6~Mpc (Freedman et al. 1994).  The present intense burst
of nuclear star formation was triggered by a tidal interaction with
M81 some $2$--5$ \times 10^8$\,yr ago (Brouillet et al. 1991; Yun, Ho
\& Lo 1994).  The active starburst region has a diameter of 500\,pc
and is defined optically by the high surface brightness regions or
clumps denoted A, C and E by O'Connell \& Mangano (1978).  An optical
imaging study of the central region of M82 with the Planetary Camera
onboard the {\it Hubble Space Telescope (HST)}\/ by O'Connell et
al. (1995) revealed the presence of over one hundred candidate SSCs
within the visible starburst. They find that the optically-bright
structures containing the clusters correspond to known sources at
X-ray, infra-red (IR) and radio wavelengths. They argue that this
cannot be accidental and suggest that regions A, C and E represent the
parts of the starburst core which are the least obscured along the
line of sight.
   
Region A is of special interest. It contains a remarkable complex of
SSCs with very high continuum and emission line surface brightnesses
(O'Connell \& Mangano 1978; O'Connell et al. 1995). The well-known
large-scale bipolar outflow or `superwind' appears to be centred on
regions A and C (Shopbell \& Bland-Hawthorn 1998; Gallagher et al., in
prep.). In a recent paper, Melo et al. (2005) have catalogued the
young stellar clusters in the starburst core of M82 using HST archive
data. They find a total of 197 optically visible clusters with 86 of
these residing in region A.

The age and evolution of the starburst in M82 have been the subject of
numerous studies. The starburst region appears to be composed of two
spatially separate stellar populations. The presence of a strong IR
continuum and large CO absorption index indicates a population of red
supergiants in the core (Rieke et al. 1993; Satyapal et al. 1997;
F\"orster Schreiber et al. 2001). The youngest and most massive stars
responsible for the strong Lyman continuum flux are located in a
$\approx 200$~pc circumnuclear arc (Achtermann \& Lacy 1995).  Using
evolutionary synthesis models, Rieke et al. (1993) and F\"orster
Schreiber et al. (2003) find that two bursts are needed to reproduce
the observational properties. The latter authors find the bursts
occurred $\approx$ 10 and 5~Myr ago with a duration of a few million
years. The first burst was particularly intense in the core and the
second occurred mainly in the circumnuclear regions.

While models of the active starburst region suggest star formation is
propagating outwards, studies of star clusters outside the central
regions reveal considerably older star formation episodes.  Gallagher
\& Smith (1999) derived ages of $60\pm20$~Myr for two SSCs M82-F and L
located 440\,pc south-west of the nucleus.  Photometric age-dating of
the extended ``fossil'' starburst region B, located 0.5--1~kpc
north-east of the nucleus, shows that the peak of the star formation
episode occurred at $\log({\rm{age/yr}})=9.0\pm0.4$ (de Grijs et
al.\ 2001, 2003, 2005).

One long outstanding question concerning starburst environments is whether the initial mass function is abnormal in the sense that less low mass stars are produced in an intense star-forming episode. M82 has been the focus of many studies aimed at settling this question. 
The mass function of the central starburst region of M82 has been estimated based on the light-to-mass ratio, but this is controversial mainly owing to the difficulty of determining extinction corrections in such a complex system.  Some studies
(e.g. O'Connell \& Mangano 1978; Telesco et al. 1991; Satyapal et
al. 1997) find a normal mass function, while others (e.g. Rieke et
al. 1993; Forster-Schreiber et al. 2003) find a deficiency of low
mass stars.  Light-to-mass ratios determined for individual SSCs
are more robust.  Smith \& Gallagher (2001) measured the dynamical mass of 
M82-F and showed that its visual
mass-to-light ratio required a mass function deficient in low mass stars, and thus it is unlikely to become an old globular cluster. McCrady, Gilbert \& Graham (2003) determined dynamical masses for two SSCs on the western edge of the active starburst region in M82, and found that one object appeared to be deficient in low mass stars.
A re-analysis of M82-F by McCrady, Graham \& Vacca (2005) based on HST
images obtained with the Advanced Camera for Surveys (ACS) and Keck/NIRSPEC spectra confirm that it has an abnormal mass function, and thus possibly a peculiar stellar mass function.

In contrast, analyses of nebular line diagnostic ratios indicate a lack of hard-UV photons in some starburst regions, implying a deficit of high mass stars (e.g. Puxley et al. 1989; Achtermann \& Lacy 1995).
In a recent study, Rigby \& Rieke (2004) suggest an alternative explanation for the low [Ne \three]/[Ne \two] ratios observed in solar metallicity-type starbursts. They propose that the hard ionizing photons from massive stars are absorbed in long-lived ultracompact H\two\ regions which are a consequence of the high interstellar densities encountered in a starburst.

The proximity of M82 ($0''.1 = 1.8$~pc) means that individual clusters
are easily resolved by {\it HST}\/ and are bright enough to obtain
spectroscopy. In this paper, we present spectroscopy obtained with the
Space Telescope Imaging Spectrograph (STIS) onboard {\it HST}\/ for five massive clusters of varying ages. The first object, which we designate as M82-A1, is a bright isolated cluster in region A. We will focus on the analysis and interpretation of this STIS dataset because it represents the first optical spectroscopy of an individual cluster and its immediate environment in the starburst core. 
We also present STIS spectroscopy for one slit position crossing both clusters F and L,
and slits covering the two brightest clusters in the fossil region B. These spectra are of lower quality than those of M82-A1 but we are able to derive spectroscopic ages.

The STIS slit position for cluster M82-A1 extends across regions A and C, close to
the nucleus and the base of the superwind, sampling the ionized gas in
the starburst core on spatial scales of a few parsecs. In a separate
paper (Smith et al., in prep.), we will discuss the nature of the
ionized component in the M82 starburst, as revealed by these optical
emission line data.
\begin{figure*}
\centering
\includegraphics[width=17.5cm,angle=0]{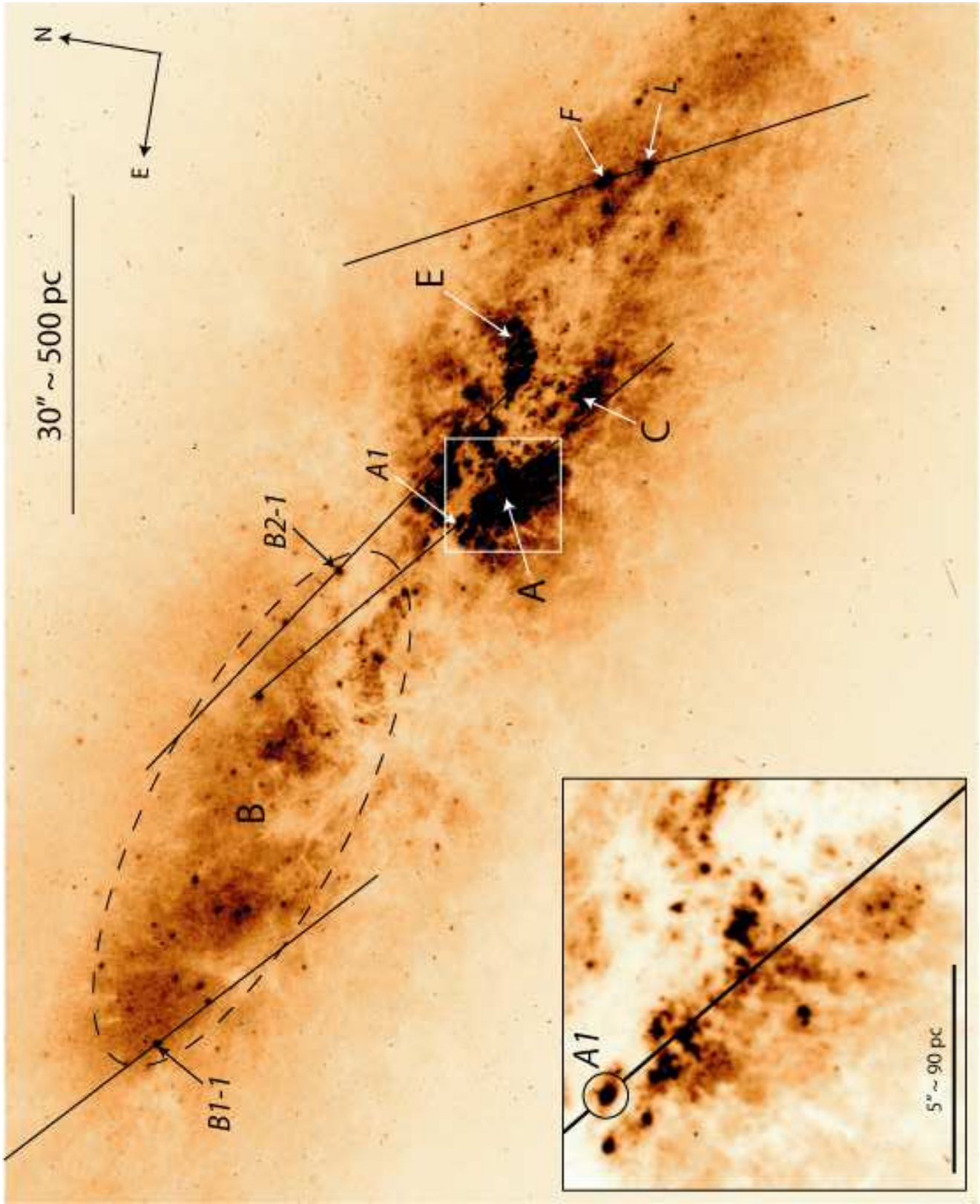}
\caption{HST$+$ACS/WFC F814W image of M82 showing the slit locations for the star clusters A1, F, L, B1-1 and B2-1. The more prominent cluster complexes (A, B, C and E) are labelled. The inset in the lower left-hand corner (corresponding to the white outline) shows the position of the STIS slit across region A and the location of cluster A1.}
\label{fig-slit}
\end{figure*} 
\section{Observations and data reduction}
\subsection {Spectra}\label{spectra}
We obtained long-slit HST/STIS spectra at four positions 
in the centre of M82 (GO 9117; P.I. O'Connell). One slit
was centred on the brightest isolated cluster in region A, which we
designate as M82-A1. Another slit sampled M82-F and M82-L, while the
two remaining pointings were centred on M82-B1-1 and M82-B2-1, within
the post-starburst region (de Grijs et al.\ 2001). 
In Fig.~\ref{fig-slit}, we show the slit positions for the five clusters
superimposed on an archive HST$+$ACS/WFC F814W image (see Sect.~\ref{images}). 
\begin{table*}
\centering
\begin{minipage}{140mm}
\caption{HST/STIS Spectroscopic Observations}
\label{stis}
\begin{tabular}{@{}llllll}
\hline
& & & \multicolumn{3}{c}{Grating}\\
\cline{4-6}
\vspace{-2.0mm} \\
&&Name&G430L&G750M&G750L\\
&&$\lambda$ range (\AA) :&2900$-$5700 &6295$-$6865 &5250$-$10300\\
Cluster&Slit&$\Delta\lambda$ (\AA\ pix$^{-1}$) :&2.73 &0.56 &4.92\\
\hline
A1&Ap=$52\arcsec\times 0\farcs1$&No. of exposures: 
&12&4&---\\
&PA$=229^{\circ}$&Date :&2002/11/27 &2002/11/26 &---\\
&Offsets=Multiple&Exposure time :&231.4m&70.8m&---\\
\\
B1-1&Ap$=52\arcsec\times 0\farcs1$&No. of exposures:
&3&2&---\\
&PA$=46^{\circ}$&Date :&2002/06/03&2002/06/03&---\\
&Offsets=none&Exposure time :&52.2m&39.m &---\\  
\\
B2-1&Ap$=52\arcsec\times 0\farcs1$ &No. of exposures: 
&3&2&---\\
&PA$=235^{\circ}$&Date :&2002/11/02&2002/11/02&---\\
&Offsets=none&Exposure time :&52.2m&39.1m&---\\
\\
F+L&Ap$=52\arcsec\times 0\farcs1$ &No. of exposures:
&---&---&4\\
&PA$=207^{\circ}$&Date :&---&---&2001/12/26\\
&Offsets=multiple&Exposure time :&---&---&163.7m\\
\hline 
\end{tabular}

A1  : $\alpha =09^{\rm h}\,55^{\rm m}\,53^{\rm s}.42$; $\delta = +69^\circ\,40'\,51''.08$ (J2000)\\
F   : $\alpha =09^{\rm h}\,55^{\rm m}\,47^{\rm s}.01$; $\delta = +69^\circ\,40'\,42''.30$ (J2000)\\
B1-1: $\alpha =09^{\rm h}\,56^{\rm m}\,03^{\rm s}.34$; $\delta = +69^\circ\,41'\,12''.20$ (J2000)\\
B2-1: $\alpha =09^{\rm h}\,55^{\rm m}\,54^{\rm s}.55$; $\delta = +69^\circ\,41'\,01''.60$ (J2000)
\end{minipage}
\end{table*}

Details of the STIS spectroscopy observations are listed in Table~\ref{stis}.  
Spectra were
taken in three different bands: the blue continuum (including the
Balmer jump), the H$\alpha$ region, and the near-infrared continuum.
Coordinates of the four clusters
initially placed at the centre of the slit are given in the footnote
to the table.  In each case, the position angle was oriented to place
other interesting regions on the slit.  In the case of A1 and F+L,
multiple exposures were made, stepped along the slit to improve the
elimination of hot pixels in the data reduction process. 

To combine the individual image sets of A1, B1-1 and B2-1, we first
registered the cosmic ray-corrected images from the {\sc
calstis} data reduction pipeline. We then flipped the cold pixels
present in the datasets to high values by inspecting intensity
histograms of the images. We next used the {\sc iraf/stsdas} cosmic
ray rejection ({\sc occrej}) task to combine all the images and
eliminate the remaining cosmic rays and hot and cold pixels.  Finally
we used the {\sc x2d} task to provide rectified absolute flux and
wavelength calibrated two-dimensional spectral images.
\begin{figure*}
\includegraphics[width=17cm]{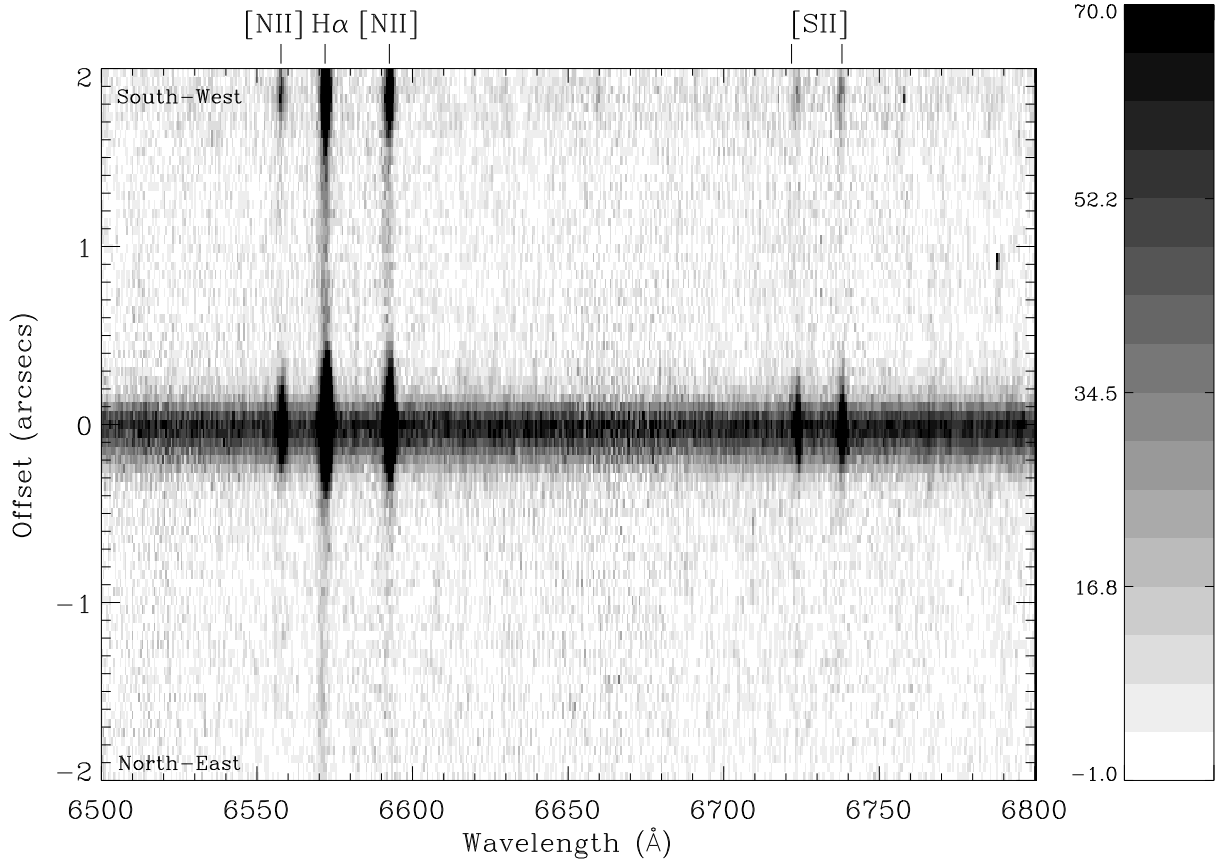}
\caption{STIS G750M two-dimensional spectral image showing the cluster M82-A1 and its compact H\two\ region over the wavelength range 6500--6800\AA.
The y-scale is in arc seconds and the slit orientation is shown. The nebular emission lines are marked and the surface brightness scale is in units of $10^{-16}$ erg cm$^{-2}$ s$^{-1}$ \AA$^{-1}$ arcsec$^{-2}$.}
\label{fig-2d-A1}
\end{figure*} 
\begin{figure*}
\includegraphics[width=17cm]{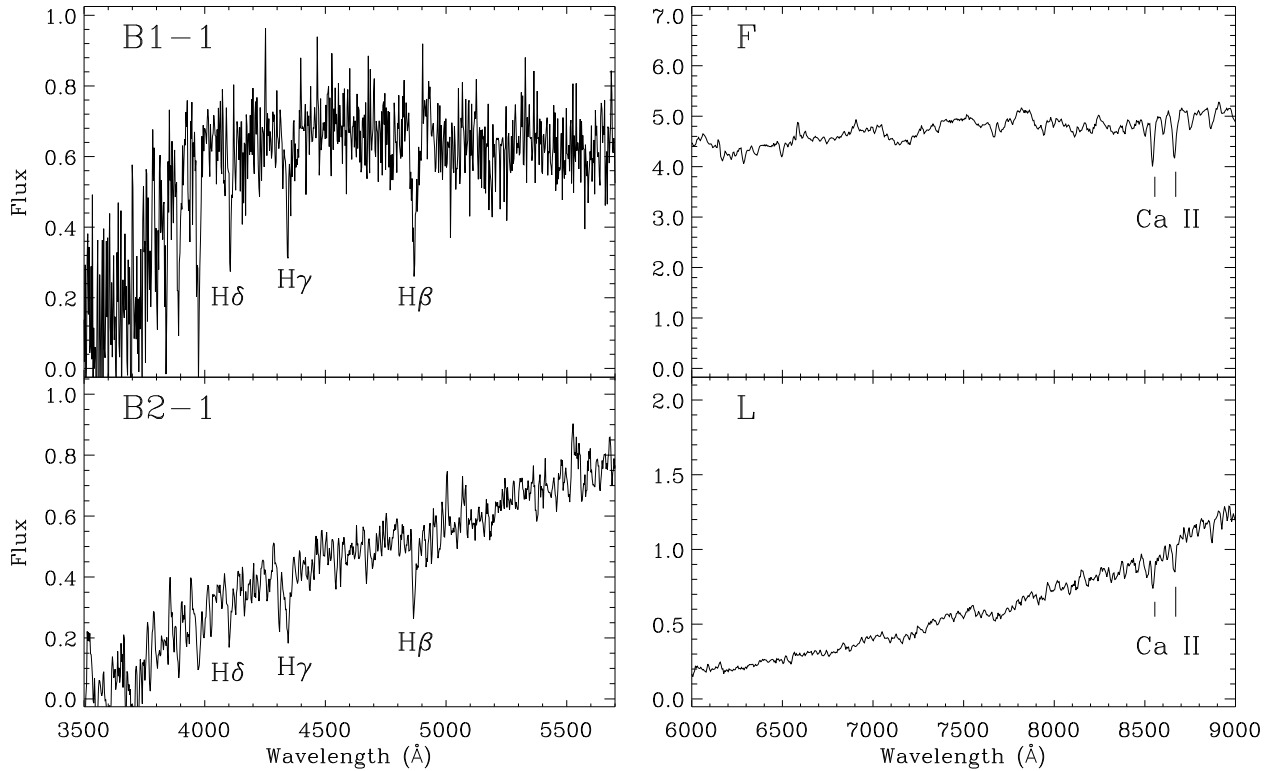}
\caption{G430L spectra of the clusters B1-1 and B2-1 (left-hand side) and G750L spectra of the clusters F and L (right-hand side). Each spectrum has been smoothed by 2\,\AA\ and the main stellar absorption features are labelled.
The flux
 is in units of $10^{-16}$ erg cm$^{-2}$ s$^{-1}$ \AA$^{-1}$.}
\label{fig-clus-sp}
\end{figure*} 

In Fig.~\ref{fig-2d-A1}, a small portion of the reduced
two-dimensional image of M82-A1 for the G750M grating is shown. The cluster is
clearly resolved and is surrounded by a compact H\two\ region. From
cuts across the STIS 2D spectra, we measure a FWHM of $\sim 6$ pixels
or $0''.3$ for M82-A1, exceeding the slit width of $0''.1$. The
spectral resolution of the G430L and G750M gratings for an extended
source is 2--3 pixels. We measure a resolution of $6.7\pm0.6$~\AA\
from Gaussian fits to the unresolved nebular H$\beta$ emission line
originating in ionized gas along the slit. We will therefore assume for this cluster
that the actual spectral resolution is 2.5~pixels or 6.7~\AA\ (G430L)
and 1.4~\AA\ (G750M grating).

To extract one-dimensional spectra, we used an extraction width of 14
pixels for clusters A1, B1-1 and B2-1, to include all the cluster
light. For A1, we performed a background subtraction using 10 pixel-wide
regions centred at $\pm25$ pixels from the centre of the
cluster. These particular regions were chosen because they contain
negligible emission from the ionized gas in region A compared to the
cluster ($<5$\% at H$\alpha$; cf. Fig~\ref{fig-2d-A1}). The exact
choice of background bins for the clusters in B1 and B2 was not
critical, so we used a prescription analogous to that for region A. No
correction for slit losses have been made.  

Clusters F+L were observed with the G750L grating. To enable the
correction of fringing effects, which are particularly prominent for
redder wavelengths, the science exposures were interspersed with CCD
flatfields.  The observations were calibrated using the {\sc calstis}
pipeline within {\sc iraf/stsdas}. After basic reduction and cosmic ray
rejection, normalized fringe flats were prepared for each exposure,
following the procedure outlined in STIS/ISR98-29. These fringe flats
were then applied to correct the fringing in the spectra. Careful
examination of the fringe-corrected spectra showed this procedure to
be quite clean and accurate. In this way, we reduced the dataset to
four, cosmic-ray cleaned and defringed, two-dimensional spectra.

Before the process of final calibration, during which a
spectrum is rectified along the wavelength and spatial axes, as 
well as converted to physical flux units, the spectra from each
of the four exposures needed to be registered, shifted and combined
into a single two-dimensional spectrum, with the associated removal 
of hot pixels. The combination was performed as follows: first, the
spatial profile of the cluster was obtained by summing over wavelength
pixels in bins which were free of contamination from both emission lines
and hot pixels lying near the position of the cluster in the spectrum. 
A separate profile was made for each exposure. 
The peak of the cluster was determined by fitting gaussians to each of
the four cluster profiles. The two-dimensional spectra were then 
shifted along the spatial axis by the amount needed to place 
the peak of the cluster exactly at the central spatial pixel of the 
spectrum. These shifted spectra were finally combined using the 
{\sc stsdas} routine {\sc mscombine}, which cleanly removes hot pixels from the final 
spectrum. 
\begin{table*}
\centering
\begin{minipage}{140mm}
\caption {HST images and derived parameters for cluster M82-A1. The errors in the effective radius and magnitude measurements are $\pm30$~mas and $\pm0.10$ mag respectively.}
\label{obs-images}
\begin{tabular}{@{}llccccccc}
\hline
Filter & Camera & Plate Scale & Exposure & FWHM & Minor/Major & $R_{\rm{eff}}$& {\sc vegamag} \\
&$+$ CCD & ($''$ pix$^{-1}$) & Time & (major axis) & Axis Ratio && Photometry & \\
&&& (s) & (mas) && (mas) & (mag)\\
\hline 
F439W & WFPC2/WF4 & 0.099 & 3300 & 140 & 0.74 & 130 &  $18.77$ \\
F439W & WFPC2/WF4 & 0.099 & 4100 & 140 & 0.71 & 140 &  $18.75$ \\
F555W & WFPC2/WF4 & 0.099 & 2500 & 170 & 0.68 & 160 &  $17.23$ \\
F555W & WFPC2/WF4 & 0.099 & 3100 & 170 & 0.65 & 160 &   $17.22$ \\
F547M & WFPC2/PC & 0.045 & 100 & 180 & 0.72 & 180 \\
F814W & WFPC2/WF4 & 0.099 & 140 & 220 & 0.66 & 210 &   $14.77$ \\
F814W & ACS/WFC &  0.049 & 120 & 210 &  0.70 & 210 & 14.91\\
F160W & NICMOS/NIC1 & 0.043 & 448 & 200 & 0.79 & 200 \\
F190N  & NICMOS/NIC2 & 0.076 & 384 & 200 & 0.71 & 190 \\
F222M & NICMOS/NIC2 & 0.076 & 448 & 160 & 0.79 & 160 \\ \\
F656N & WFPC2/PC & 0.045 & 1000 & 290 & 0.57 & 260 & 8.54E$-14^{1}$\\
\hline
\end{tabular}

$^{1}$Total H$\alpha$ flux in units of ergs\,cm$^{-2}$\,s$^{-1}$
\end{minipage}
\end{table*}

Since the clusters are marginally resolved with the $0.1''$ aperture of the slit,
aperture illumination effects can be substantial. Therefore, we used
the {\sc stsdas} routine {\sc x1d}, part of the {\sc calstis} ensemble, to extract the
final one-dimensional spectrum of clusters F and L within 11 pixel
wide bins, as well as adjacent galaxy backgrounds. Using X1D ensures
that wavelength dependent aperture illumination effects are taken into
account during flux calibration of the spectrum. The backgrounds were
then subtracted from the cluster extractions to obtain spectra of pure
cluster light.  For cluster F, the background accounts for no more
that $10$\% of the light in the cluster extraction, but this fraction
is higher for the lower luminosity cluster L, especially in the blue
end of the spectrum, since cluster L is heavily reddened by
intervening dust.

In Fig.~\ref{fig-clus-sp},we show the final reduced G430L spectra for clusters B1-1 and B2-1, and the G750L spectra for clusters F and L. The spectra for cluster A1 are presented in Sect.~\ref{sect_spectra}. 

\subsection{Images}\label{images}
In this section we present HST archival images for cluster M82-A1 and measure its size and perform photometry. HST images and measurements have been presented  by de Grijs et al. (2001) for clusters B1-1 and B2-1. In their nomenclature, B1-1 corresponds to No. 28 in their table 2, and B2-1 to No. 12 in their table 3. The $V$ magnitudes of B1-1 and B2-1 are given as 17.93 and 17.90 mag respectively. Smith \& Gallagher (2001) present HST images of clusters F and L 
obtained with the Wide Field and Planetary Camera 2 (WFPC2) while McCrady et al. (2005) present images of cluster F obtained with the High Resolution Channel (HRC) of the ACS instrument.

We obtained HST images taken with the WFPC2, ACS and NICMOS cameras
for the central region of M82 from the HST archive; the details of
these images are given in Table~\ref{obs-images}.  The broad-band
WFPC2 F439W, F555W and F814W images from the GO 7446 programme (PI
O'Connell) are centred on the fossil starburst regions B1 and B2 (see
de Grijs et al. 2001) and region A is on the WF4 CCD. We also obtained
narrow-band images in the F656N and F547M filters which are part of
the GO 6826 programme (PI Shopbell) on the superwind of M82.  We used
the F547M filter for continuum subtraction in the F656N filter.  These
images are discussed in more detail in Gallagher et al. (in prep.).
The single ACS F814W image was obtained as part of the GO 9788
snapshot programme (PI Ho) and is shown in Fig.~\ref{fig-slit}. The NICMOS images from the GO 7218
programme (PI Rieke) are described in Alonso-Herrero et al. (2003).

Cosmic rays were removed from the pipeline-reduced images using the Laplacian cosmic ray identification routine of van Dokkum (2001). The  individual images were then checked for saturation and averaged. Photometry and radius measurements of M82-A1 are complicated by the varying background of emission and absorption by gas and dust in the starburst core. The images also show that the cluster is quite elliptical (Fig.~\ref{fig-slit}).

To measure the size of M82-A1, we used the {\sc ishape} algorithm (Larsen 1999) together with the {\sc tinytim} package (Krist 2004) to correct for the point-spread function (PSF) for the WFPC2 and NICMOS cameras.  For the ACS F814W image, we used an empirical PSF generated from point sources in the field of 47 Tuc (Mora, priv. comm.).
First, we determined the best fitting radius for the profile fits  by examining fits to the F547M image using a Moffat function with a power index of 1.5. We found that a
radius of 0.5 arcsec was optimal since this was sufficiently large but excluded light from other nearby sources.
We then experimented with different profile types and chose a Moffat function with a power index of 1.5 since this gave the lowest residuals in the overall fit. We also
tried varying the power-law index of the Moffat profile but found that a value of
1.5 consistently gave the best fit. Elson, Fall \& Freeman (1987) determined that profiles of this type provide good fits to LMC star clusters while de Grijs et al. (2001) found similar results for the profiles of the brightest clusters in region B of M82, including B1-1 and B2-1.

In Table~\ref{obs-images}, we list the FWHM of the major axis
and the minor/major axis ratio, as estimated by {\sc ishape} for each image. 
We also give the effective or half-light radius $R_{\rm{eff}}$ 
where we have used the expressions given by 
Larsen (2004) to convert the FWHM of an elliptical Moffat profile fit to a
half-light radius. We find very consistent results
for the WF4 and ACS F814W measurements given the different pixel scales and
PSFs. For comparison, as detailed in the instrument handbooks, the PSFs of the WF4, PC and ACS/WFC chips are $\approx 80, 130$ and 80 mas respectively.
To estimate appropriate errors, we fitted the F547M and F814W/ACS images
with a number of different profile types. We find that, in terms of the residuals,
King profiles with different concentration parameters also give acceptable fits. From the
scatter in the resulting  $R_{\rm{eff}}$ values, we find that the likely error in each
measurement is $\pm30$~mas.
Taking a straightforward mean of the radii measured in the continuum filters, 
we find  $R_{\rm{eff}}= 170\pm 30$~mas or $3.0\pm0.5$~pc
for M82-A1. We derive a mean minor/major axis ratio of $0.72\pm0.05$ from the same images. For comparison, McCrady et al. (2003) have measured $R_{\rm{eff}}$ for several clusters (including M82-A1 which is MGG-a1 in their designation) from NICMOS NIC2 F160W images. From spherical King profile fits, they find $R_{\rm{eff}}=119\pm14$~mas which is smaller than our own more precise measurements which allow for ellipticity. It is apparent from Table~\ref{obs-images} that the radius of M82-A1 decreases towards shorter wavelengths; we discuss this effect in Sect.~\ref{sect-radius}.
				   
Measurements of the size and total flux of the
compact H\two\ region surrounding M82-A1, based on H$\alpha$ emission measured in the F656N filter, are also given in Table~\ref{obs-images}. It is highly elliptical with a major axis FWHM value of 290 mas and an axis ratio of 0.57, giving a half-light
radius $R_{\rm{eff}}$ of $260\pm30$~mas or $4.5\pm0.5$~pc. Thus the H\two\ region is only 1.5 times larger than the cluster.

In Table~\ref{obs-images}, we list the magnitudes of M82-A1 for the wide-band filters in the {\sc vegamag} system. M82-A1 is superimposed on a varying background. 
We tried various approaches to subtracting the background and found that the most
robust and simplest method was to use a flat plane. The level was calculated from
an average of measurements, taken at four locations around the cluster M82-A1, that were deemed to contain ``representative'' background values. The photometry of M82-A1 was performed using the {\sc phot} task within the NOAO {\sc digiphot} package. It was decided that a 7 pixel radius aperture gave the best results, considering the morphology of M82-A1 is such that it has a protrusion extending north-westwards of the cluster.
The total counts measured in each aperture were corrected for CCD charge transfer effects using the formulation of Dolphin (2002). They were then converted to magnitudes using the zeropoints provided for the {\sc vegamag} system in the instrument handbooks. We estimate the measurements are accurate to $\pm0.10$ mag because of the uncertainties in the varying background. This level of accuracy was also verified with two sets of independent measurements.

\begin{figure}
\psfig{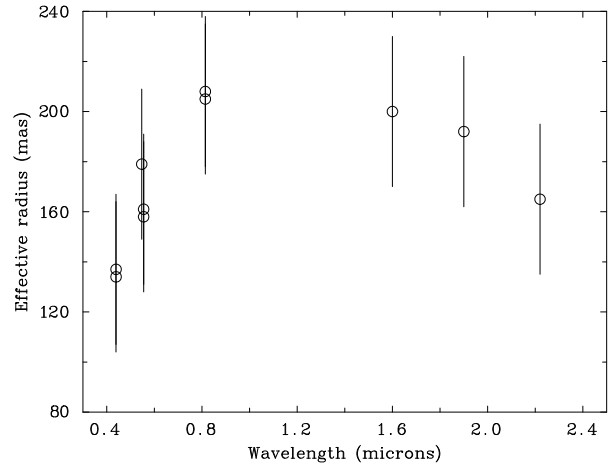}
\caption{Measured effective radii (in milli-arcsec) of M82-A1 plotted as a function of wavelength using the data in Table~\ref{obs-images}. The errors bars represent the measurement uncertainties of $\pm30$ mas.}
\label{fig-radii}
\end{figure} 

\section{Cluster M82-A1}
\subsection{The change of radius with wavelength}\label{sect-radius}
Before presenting and discussing the STIS spectra of M82-A1, we first discuss possible reasons for the decrease in radius of the cluster at progressively shorter wavelengths.
In Fig.~\ref{fig-radii} we plot
$R_{\rm{eff}}$ against wavelength for M82-A1.  If we exclude the
NICMOS measurements in Table~\ref{obs-images} (because the diffraction limit is close to the
radius of M82-A1), then the cluster radius appears to decrease at
successively shorter wavelengths; the radius is 30\% smaller in the
F439W filter compared to the F814W filter.

McCrady et al. (2005) find that the radius of cluster M82-F
decreases with increasing wavelength, and they interpret this as being
due to mass segregation, with the more centrally concentrated stars being the most massive cooler evolved stars. For M82-A1, we see the opposite effect with the radius deceasing towards blue wavelengths.

It is not clear what is causing this decrease in radius. At the young age of this cluster, most of the detected light should originate from stars with roughly the same mass.
We find from the output of the spectral synthesis code Starburst99 
(see Sect.~\ref{age}) that this is indeed the case, and that any change in colour with mass will be shallow for a mass segregated cluster. This leads us to conclude that it is difficult to produce a radial gradient due to mass segregation. 

\begin{table*}
\centering
\begin{minipage}{140mm}
\caption {Absorption and emission line measurements for M82-A1. Velocities are in the heliocentric frame of reference.}
\label{spectra-meas}
\begin{tabular}{@{}lcrrrl}
\hline
Line & $\lambda_{\rm {vac}}$ & Velocity & FWHM & Equivalent & Comments\\ 
     & (\AA)                    & (km\ s$^{-1}$) & (km\ s$^{-1}$) 
     & Width (\AA) \\
\hline
\noalign {\it G430L spectrum}\\
{[}O\two{]} & 3728.48 & 448 & & $-2.9\pm0.6$ & blended doublet\\
H8 abs & 3890.15 & 171 & &$3.2\pm0.5$ & blended with He\one ?\\
Ca\two\ K & 3934.78 & 205 & & $3.8\pm0.6$& Interstellar\\
Ca\two\ H & 3969.59 & 415 & & & Interstellar; blended with H$\epsilon$\\
H$\epsilon$ abs & 3971.19 & 294 & & $3.9\pm0.7$ & blended with Ca\two\ H\\
H$\delta$ abs & 4102.89 & 335 && $3.0\pm0.7$\\
H$\gamma$ em & 4341.68 & 332 \\
H$\gamma$ abs & 4341.68 & 327 && $2.0\pm0.9$\\
H$\beta$ em & 4862.68 & 298 && $-1.8\pm0.2$\\
{[}O\three{]} & 4960.30 & 343  && $-1.0\pm0.1$\\
{[}O\three{]} & 5008.24 & 319  & & $-1.8\pm0.2$\\
\\
\noalign {\it G750M spectrum}\\
{[}N\two{]} & 6549.85 & 322 & 93 & $-4.8\pm0.2$\\
H$\alpha$ & 6564.61 & 319 & 95 & $-27.6\pm0.3$\\
{[}N\two{]} & 6585.28 & 318 & 96 & $-15.5\pm0.2$\\
{[}S\two{]} & 6718.29 & 319 & 77 & $-2.3\pm0.2$\\
{[}S\two{]} & 6732.67 & 322 & 84 & $-3.1\pm0.2$\\
\hline
\end{tabular}\\
\end{minipage}
\end{table*}

We now consider whether the compact H\two\ region surrounding the cluster could affect the radius measurements.  It is unlikely that contamination from nebular emission lines in the broad-band filters is causing the radius variations because the emission lines are weak (see Figs.~\ref{fig-sp-blue} and ~\ref{fig-sp-red}) and the F814W filter does not contain any significant nebular emission. The presence of dust in the H\two\ region may, however, produce a decrease in radius towards bluer wavelengths.
A diffuse foreground screen of dust will change the brightness but, since it will dim the cluster uniformly, it will not affect the radius or shape as a function of wavelength. However, if dust is concentrated in the  H\two\ region, then the cluster would maintain a roughly constant shape (as measured), yet be dimmer and smaller with decreasing wavelength because of the greater obscuration at shorter wavelengths.
We therefore conclude that the most probable explanation for 
the decrease in radius with decreasing wavelength is the effect of dust in the surrounding compact H\two\ region.  To resolve this conclusively, ACS/HRC observations of M82-A1 are required. Dynamical models addressing the colour gradients in young mass segregated clusters and a check for cool supergiants that could support colour gradients are also needed.

\subsection {Spectra}\label{sect_spectra}
\begin{figure*}
\psfig{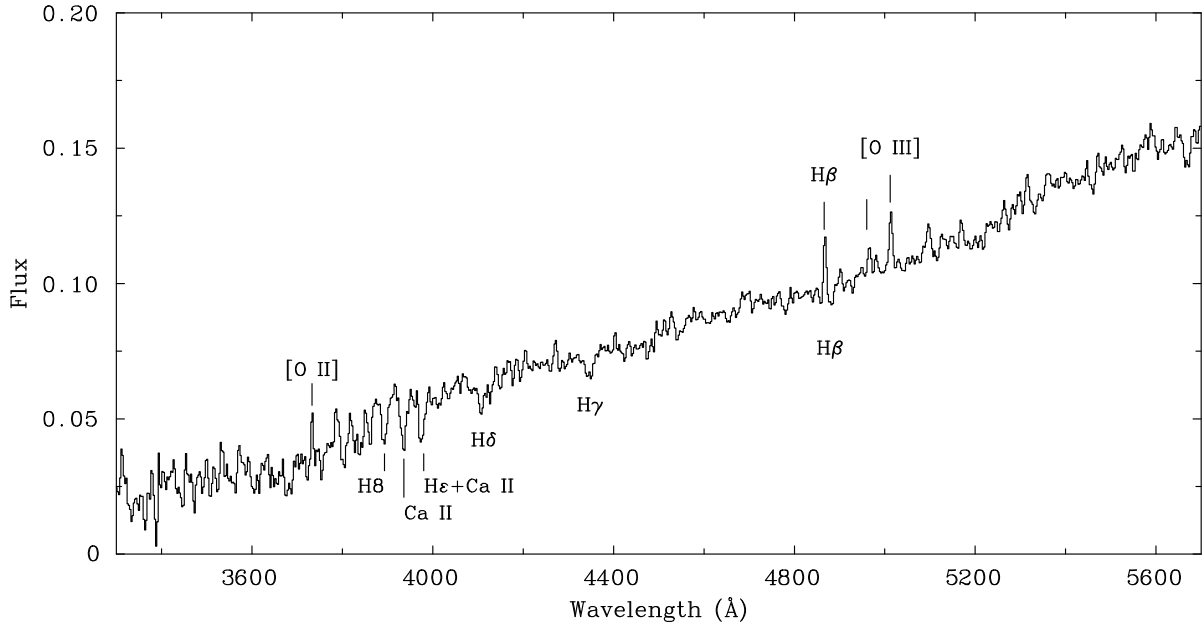}
\caption{G430L spectrum of the cluster M82-A1 and its H\two\ region smoothed by 2~\AA.
The nebular emission and cluster absorption lines are marked above and below
the spectrum respectively.
The flux is in units of $10^{-15}$~erg~s$^{-1}$~cm$^{-2}$~\AA$^{-1}$}
\label{fig-sp-blue}
\end{figure*}

\begin{figure*}
\psfig{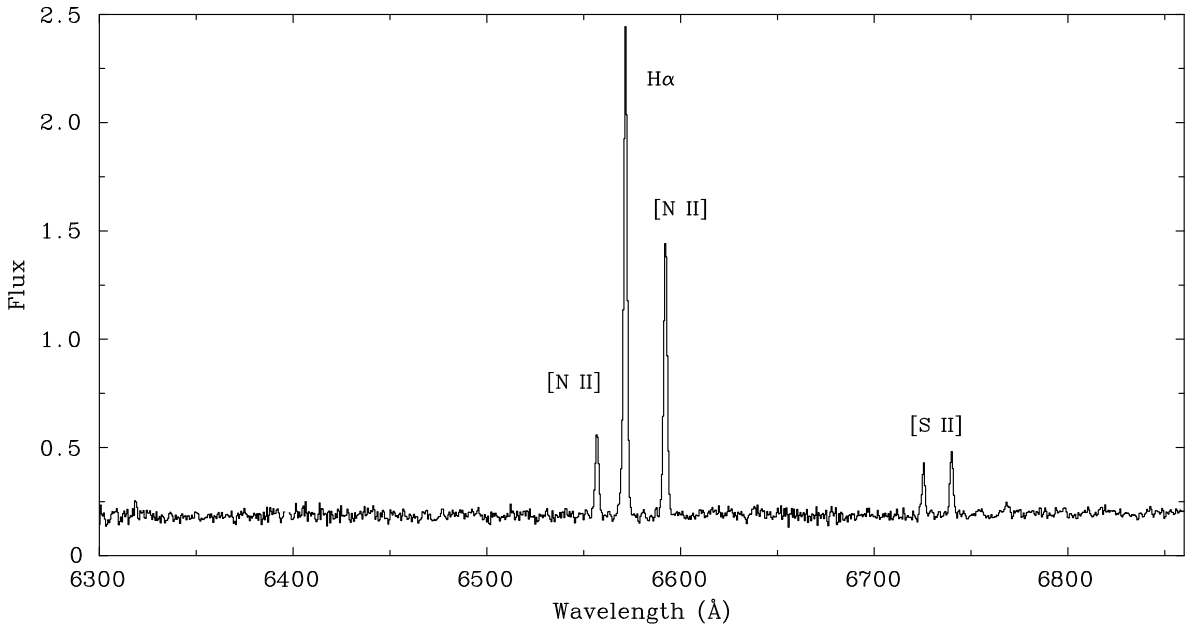}
\caption{STIS G750M spectrum of the cluster M82-A1 and its H\two\ region.
The nebular emission lines from the surrounding H\two\ region are marked.
The flux is in units of $10^{-15}$~erg~s$^{-1}$~cm$^{-2}$~\AA$^{-1}$}
\label{fig-sp-red}
\end{figure*}

In Figs.~\ref{fig-sp-blue} and \ref{fig-sp-red}, we show the flux-calibrated spectra of M82-A1 for the G430L and G750M gratings. The major line identifications,
heliocentric velocities, FWHMs (for the G750M spectrum only) and 
equivalent widths are presented in 
Table~\ref{spectra-meas} for the absorption and emission features. The equivalent widths were measured by defining continuum windows in the spectra and the errors represent $1\sigma$ uncertainties based on the error in the continuum.

\subsubsection{Absorption Lines}
The G430L spectrum is dominated by Balmer absorption lines from the cluster, and
a weak Balmer jump at 3650~\AA, indicating a young age.  Strong interstellar Ca\two\ absorption lines are also present. No stellar absorption lines are detected in the G750M spectrum.
The mean heliocentric radial velocity of the cluster as measured from the Balmer absorption lines is $319\pm22$~\kms. The velocity of the interstellar Ca\two\ K line is $+205$~\kms. This is very close to the systemic velocity of M82 of $+200$~\kms 
(Achtermann \& Lacy 1995). The cold interstellar gas therefore appears to be at rest for this line of sight near the nucleus of M82.

\subsubsection{Emission Lines}
A number of emission lines arising from the compact H\two\ region surrounding
M82-A1 are apparent in the STIS spectra. We detect nebular emission lines of
H$\alpha$, H$\beta$, H$\gamma$, [O\two] $\lambda3727, 3730$, [O\three]
$\lambda4960, 5008$, [N\two] $\lambda 6550, 6585$ and [S\two] $\lambda 
6718, 6733$. The low [S\two]/H$\alpha$ ratio of 0.21 is consistent with 
photoionization rather than shock heating. In Fig.~\ref{fig-sp-red}, 
the red member of the [S\two] doublet is the strongest
of the two lines, indicating that the H\two\ region has a high density.
In Section~\ref{para-hii} we derive the electron density and oxygen abundance.

The comparatively high resolution of the G750M spectrum allows us to measure accurate velocities and FWHMs for the H\two\ region, as given in Table 2. We find a mean heliocentric velocity for the H$\alpha$, [N\two] and [S\two] nebular lines of $320\pm2$ \kms, in excellent agreement with the measured value of the cluster radial
velocity. 
The lines have an observed FWHM of $89\pm8$~\kms\ and are clearly resolved since
the  measured resolution is 63~\kms; the intrinsic mean FWHM is then $62\pm11$~\kms.
This gives a maximum expansion velocity for the H\two\ region of $\sim 30$~\kms.
\subsection{Cluster parameters}\label{age}
The starburst core of M82 suffers from a high and non-uniform
extinction. For example, Satyapal et al. (1995) find values for the
visual extinction $A_{\rm V}$ ranging from 2--12 mag from near-IR
nebular line ratios, assuming a foreground screen model.  For the
whole M82-A complex, centred about 6\arcsec\ SW of M82-A1 (see
Figs.~\ref{fig-slit}a and \ref{fig-slit}b), Marcum \& O'Connell (1996)
estimated that $E(B-V) \sim 0.60$ arises in a foreground dust screen
with an additional $\tau$(5500 \AA) $\sim 0.5$ arising from internal
dust.  As can be seen from Figs.~\ref{fig-slit} and \ref{fig-2d-A1},
the STIS slit crosses some areas of complete obscuration at optical
wavelengths. The cluster M82-A1 sits close to the conspicuous dust
lane that bisects the galactic disk; to the east of this cluster (or
lower pixel number in Fig.~\ref{fig-2d-A1}), we detect little emission
due to the very high extinction.
\begin{figure*}
\psfig{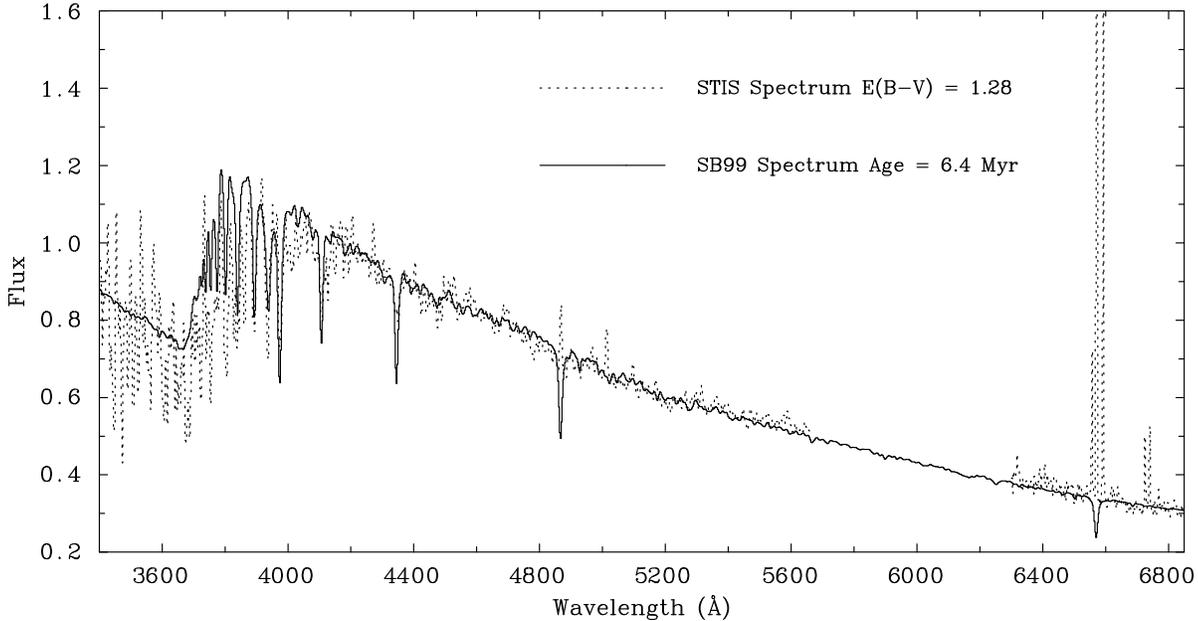}
\caption{Comparison of the de-reddened ($E(B-V=1.28)$) STIS spectrum of M82-A1 
with a synthetic Starburst99 (SB99) spectrum for an age of 6.4 Myr. For this
combination of age and reddening, the depth of the Balmer jump and the overall
SED are well-fitted.
The flux is in units of $10^{-14}$~erg~s$^{-1}$~cm$^{-2}$~\AA$^{-1}$}
\label{fig-age}
\end{figure*} 
\begin{figure}
\psfig{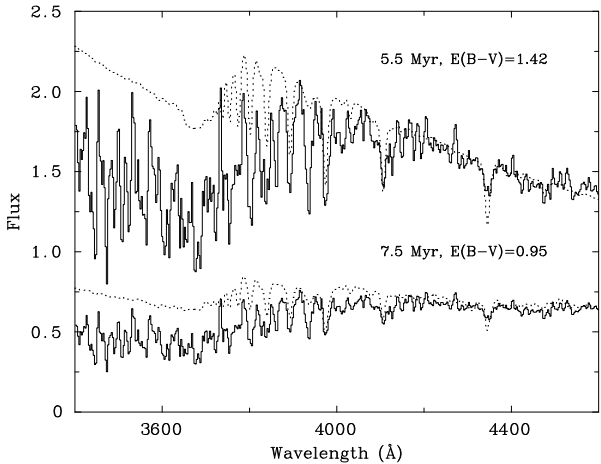}
\caption{Comparison of the de-reddened STIS spectrum of M82-A1 
with synthetic SB99 spectra for ages of 5.5 Myr ($E(B-V)=1.42$) and 7.5 Myr
($E(B-V)=0.95$) demonstrating that the depth of the Balmer jump cannot be fitted at these ages. The flux is in units of $10^{-14}$~erg~s$^{-1}$~cm$^{-2}$~\AA$^{-1}$ and
the spectra for the 7.5 Myr comparison have been multiplied by a factor of 2.5.}
\label{fig-out}
\end{figure} 

Ages and reddening for clusters less than 10 Myr old are usually
determined from spectroscopy using the UV wavelength region because
the spectral energy distribution (SED) is largely independent of age
and the stellar wind features are sensitive age indicators (e.g.
Tremonti et al. 2001; Chandar et al. 2004).  However, given sufficient
spectral resolution, the short-wavelength optical band (especially the
continuum structure and strong absorption lines in the Balmer jump
region between 3500 and 4100 \AA) should provide a comparable hold on
age and reddening.  Gallagher \& Smith (1999) derived the age of
cluster M82-F by comparing the observed spectrum in the region of the
Balmer jump with theoretical model cluster spectra.
Here, we use recent high-resolution population synthesis models to fit
the combined spectrum of M82-A1.

First, we binned the G750M STIS spectrum to match the G430L spectrum
and merged the two spectra.  We fit the
combined spectrum using v5.0 of the spectral synthesis code
Starburst99 (SB99: Leitherer et al.  1999; Gonz\' alez Delgado et al.
2005) which includes a new synthetic high resolution (0.3\,\AA)
spectral library from 3000--7000\,\AA\ containing non-LTE models for
OB stars (Martins et al. 2005).  We generated a series of SB99 models
for different ages (2--10 Myr), assuming solar metallicity (see
Sect.~\ref{para-hii}), an instantaneous burst with a Kroupa (2001)
IMF, lower and upper masses of 0.1 and 100~\Msun, and the enhanced
mass loss Geneva tracks.  

The resulting high resolution synthetic spectra were smoothed, binned, and velocity shifted to match the G430L spectrum of M82-A1. We de-reddened the observed combined spectrum using a foreground dust screen, employing the Galactic reddening law from Howarth (1983) and $R= A(V)/E(B-V)=3.1$, and a range of $E(B-V)$ values from typically 0.9--1.8 mag. These de-reddened spectra were then normalised to the synthetic spectra at 6800 \AA\ and the quality of the fit for each possible combination of age and $E(B-V)$ was judged by the $\chi^2$ statistic, with
highest weight given to the spectral region 3700--4400 \AA. Typically, we find that we can match the model continuum longward of the Balmer jump from 3700--6800 \AA\ by varying $E(B-V)$ until a good fit is obtained; this then provides a value of the reddening for each model as a function of age. The quality of the fit of the model spectrum to the observed de-reddened spectrum below the Balmer jump then determines the age.

Using this approach, we find that the STIS spectra of M82-A1 can only
be matched by the SB99 synthetic spectra for a relatively narrow age
range of 5--7\, Myr and $E(B-V)=1.5$--1.2 mag. We checked the SB99
results with a comparable grid taken from the independent synthesis
models of Bruzual \& Charlot (2003; BC03), which incorporate a lower
resolution (3 \AA) but more realistic empirical spectral library.  The
fits are quite similar to those of SB99, with a slightly larger age
range of 4--7\, Myr.  From ground-based spectrophotometry, Marcum \&
O'Connell (1996) obtained an average age of $\sim 5$ Myr for the
entire M82-A complex, in good agreement with these estimates for
M82-A1.

In Fig.~\ref{fig-age} we show the SB99 fit for the best fitting SB99
model, at an age of 6.4 Myr and $E(B-V)=1.28$ mag.  The overall fit to
the continuum is good, and it can be seen that the upper members of
the Balmer series are also fairly well-fitted (e.g. H8 which has
minimal nebular line emission).  This model, which has the largest
Balmer jump in either model grid, slightly overestimates the observed
flux in the Balmer continuum below 3600 \AA. The discrepancy could be
caused by internal extinction within M82-A1 or perhaps a small error
in the STIS calibration here.  Alternatively, there could be
weaknesses in the models.  Adjacent models which fit the continuum above the Balmer jump by varying the $E(B-V)$ parameter provide poorer
short-wavelength fits below the Balmer jump, as illustrated in Fig.~\ref{fig-out}.  Abrupt
changes in the Balmer jump and the color of the continuum in the SB99
models for ages of 6--10 Myr are probably caused by intermediate
temperature class I-II stars, and it is not clear that these are
being accurately modelled (Gonz\' alez Delgado et al.
2005).

We can now employ two independent indicators for reddening and age to
determine if they are consistent with those found from the spectral
analysis. First, we use the nebular H\one\ emission lines to measure
the interstellar extinction by comparing the observed
H$\alpha$/H$\beta$ and H$\gamma$/H$\beta$ ratios with the intrinsic
case B values of Hummer 
\& Storey (1987). We correct for stellar absorption by directly subtracting 
the reddened synthetic SB99 spectrum from the observed STIS spectrum using
the reddening values we have determined for the best-fit age range.
We find $E(B-V)=1.38\pm0.15$ mag, in good agreement 
with dereddening the cluster energy distribution. This method of directly subtracting
the synthetic spectrum to recover the intrinsic nebular line fluxes has the
advantage that we can recover the H$\gamma$ emission line and obtain a more 
accurate measurement for the H$\beta$ line flux.
We also compare the observed $(B-V)$ and $(V-I)$ colours of M82-A1 
with the predicted colours from SB99 for ages of 6 and 7 Myr and find 
$E(B-V)=1.54\pm0.26$ in reasonable agreement with the above.
For the reddening towards M82-A1, we will adopt the mean value given by the
cluster SED and the nebular lines of $E(B-V)=1.35\pm0.15$ mag.

Our independent age indicator is the strength of the observed
equivalent width of the nebular H$\alpha$ emission (Table~\ref
{spectra-meas}) when compared with the predicted values from
Starburst99.  The observed equivalent width of $32\pm2$~\AA\
(corrected for stellar absorption) gives an upper limit to the age of
6.7 Myr.  This value is consistent with the age determined from the
Balmer jump, and suggests that few ionizing photons escape from the
H\two\ region. We thus conclude that the age of M82-A1 is  $6.4\pm0.5$
Myr.

Using the above values, we derive $M_{V}= -14.84\pm0.38$ and 
$\log L_{V}/L_{\odot}=7.87\pm0.15$. The photometric mass is very
dependent on the form of the IMF. For a Kroupa (2001) IMF and a
Salpeter IMF (with a lower mass limit of 0.1 M$_{\odot}$), we derive
masses of  $6.6^{+0.3}_{-0.2}\times10^{5}$  and
$1.3^{+0.5}_{-0.4}\times10^{6}$ M$_{\odot}$ respectively.
The derived parameters of M82-A1 are given in Table~\ref{param}.
\begin{table}
\caption {Summary of derived parameters for cluster M82-A1 and its H\two\ region.}
\label{param}
\begin{tabular}{@{}lr@{}c@{}l}
\hline
Parameter && \multicolumn{2}{l}{Value} \\
\hline 
{\bf \ \ \ Cluster}\\
Radial velocity $V_{\rm r}$ & +320 & $\pm$ & 20 km s$^{-1}$ \\
F555W & $17.23$ & $\pm$ & $0.10$ mag\\
Half-light radius $R_{\rm{eff}}$ & 3.0& $\pm$ & 0.5 pc \\
$E(B-V)$ & 1.35 & $\pm$ & 0.15 mag\\
$M_{V}$ & $-14.84$ & $\pm$ & 0.38 mag \\
Log $L_{V}$/$L_{\odot}$ & 7.87 & $\pm$ & 0.15 \\
Age & 6.4 & $\pm$ & 0.5 Myr \\
Mass $M$ & 6.6 & $^{+}_{-}$ &$^{0.3}_{0.2}\times10^{5}$ M$_{\odot}$$^{1}$ \\
&1.3 & $^{+}_{-}$ &$^{0.5}_{0.4}\times10^{6}$ M$_{\odot}$$^{2}$ \\
Photon Luminosity  & 7.5 & $\pm$ & $3.0 \times 10^{50}$ s$^{-1}$\\
in Lyman continuum $Q_{0}$\\
\\
{\bf \ \ \ H\two\ Region}\\
Half-light radius $R_{\rm{eff}}$ & 4.5& $\pm$ & 0.5 pc  \\
Electron density $N_{\rm e}$ & $1800$ &$^{+}_{-}$& $^{340}_{280}$ \cm3 \\
$\log O/H + 12$ && $\sim$ & 8.8\\
Ionization  & $-2.24$ & $\pm$ & $0.18$ \\
Parameter $\log U$\\
Ionized Mass && $ >$ & 5\,000 M$_{\odot}$ \\
\hline
\end{tabular}

$^{1}$ using a Kroupa (2001) IMF; $^{2}$ using a Salpeter IMF.
\end{table}

\subsection{H\two\ region parameters}\label{para-hii}
We now turn our attention to the properties of the H\two\ region that surrounds
the cluster M82-A1. We have already established that it is compact with
a half-light radius of $4.5\pm0.5$~pc.
From Gaussian fits to the [S \two] emission line profiles, we derive an electron density
\Ne$=1800^{+340}_{-280}$~\cm3, assuming a standard electron temperature 
value of \Te$=10^{4}$~K. The H\two\ region surrounding M82-A1 can therefore
be classified as a high pressure ($P/k =1$--$2 \times 10^{7}$~\cm3\,K, considering the likely range in \Te) compact 
H\two\ region.  
We are also able to derive $\Ne$ for the diffuse gas along the STIS slit in
region A and find a mean value of $\sim 1000$~\cm3, suggesting that the
ionized gas in region A  is at high pressure, at least the region
probed by the optical [S\two] lines. O'Connell \& Mangano (1978) derived a
similar value of \Ne$=1800$~\cm3 for regions A and C from the [S\two] ratio.

Our $\Ne$ measurements are far higher than those of F\"orster Schreiber et al. 
(2001) who measured \Ne\ in the starburst core using observations obtained with
the Short Wavelength Spectrograph (SWS) onboard the {\it Infrared
Space Observatory}. The SWS aperture covers an area of $14'' \times
20''$ and they find that the line ratios of the fine structure
lines of [S\three], [Ne\three] and [Ar\three] all lie at the low
density limit; they adopt \Ne$=300$\,\cm3.  
The large difference in density between the STIS and ISO measurements
may be due to sampling different environments. We consider this in more
detail in the next section when we discuss the evolution of the M82 starburst
as revealed by the study of F\"orster Schreiber et al.  (2003).
\begin{figure*}
\psfig{file=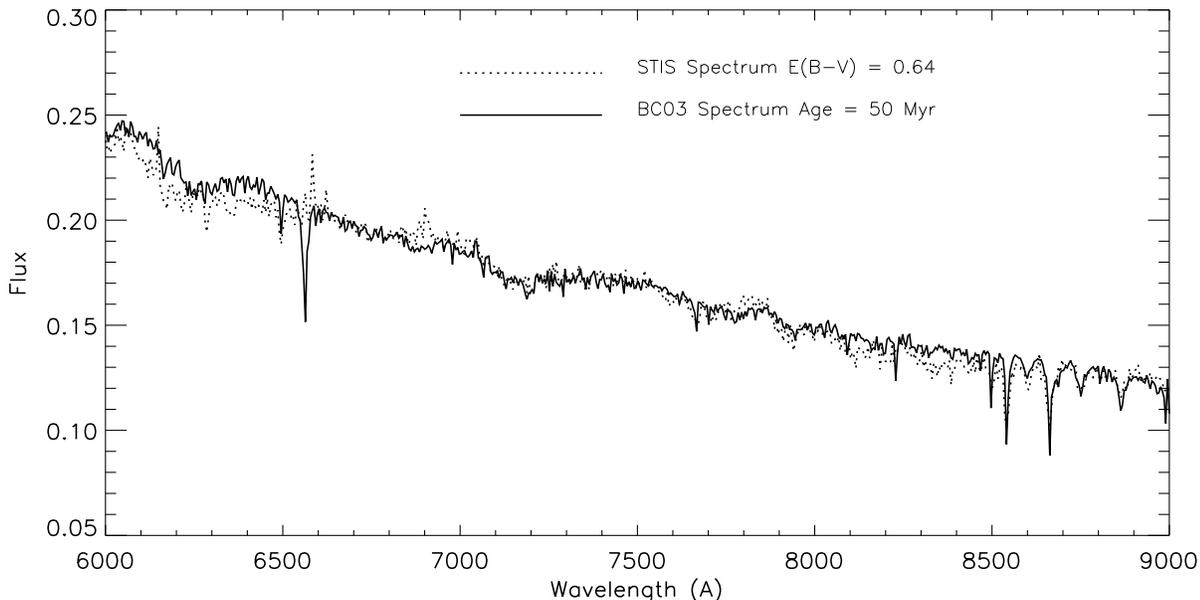,width=18cm,bbllx=20pt,bblly=20pt,bburx=370pt,bbury=750pt,angle=90}
\caption{Comparison of the de-reddened ($E(B-V)=0.64$) STIS spectrum of M82-F 
with a synthetic BC03 spectrum for an age of 50 Myr.
The flux is in units of $10^{-14}$~erg~s$^{-1}$~cm$^{-2}$~\AA$^{-1}$}
\label{fig-F}
\end{figure*} 

We can obtain an approximate oxygen abundance for the H\two\ region using
techniques that rely on the relative
strengths of the strong emission lines when \Te\ is unknown. We will use two 
recent empirical methods. First we
follow Kewley \& Dopita (2002) and use their equation (7) to derive $O/H$
from the de-reddened [N\two]/[O\two] ratio of 2.2. We find $\log O/H+12=9.23$
based on a solar O abundance of $\log O/H+12=8.9$. Thus the O abundance using
this method is $\sim 2\times$ solar. 
Next, we use the method of Pettini \&
Pagel (2004) which is based on the ratios of 
[N\two]$\lambda6583$/H$\alpha$  ($N2$) and 
[O\three] $\lambda5007$/H$\beta$ to [N\two]$\lambda6583$/H$\alpha$
($O3N2$) and is insensitive to reddening. Using the relations given by
Pettini \& Pagel, we derive $12+\log O/H = 8.76$ ($N2=-0.25$) and 8.76
($O3N2=-0.087$) based on a solar O abundance
of $8.69$. Thus this method gives an O abundance of $\sim 1.2\times$ solar.
Finally, if we take the derived $\Ne$ and adopt a value of \Te$=10^{4}$~K,
we derive $12+\log O/H = 8.77$. Although these methods are approximate,
they indicate that the oxygen abundance of the H\two\ region surrounding
M82-A1 is close to solar, and that our assumption that \Te$=10^{4}$~K is reasonable.

For comparison, F\"orster Schreiber et al. (2001)
derive Ne, Ar and S abundances from their SWS dataset. They find that
the abundances of Ne and Ar are slightly above solar and S is 0.25
solar.  In a recent paper, Origila et al. (2004) have obtained abundances
for red supergiant stars in the nuclear region of M82 and find a solar
oxygen abundance.

We can also derive the ionized mass of the H\two\ region from the observed total H$\alpha$ flux listed in Table~\ref{obs-images}. Using the adopted value for $E(B-V)$, a distance of 3.6 Mpc, \Te$=10^{4}$ K, and \Ne$=1800$ \cm3, we find an ionized mass of
$>5\,000$ M$_{\odot}$. This value is a lower limit because the close agreement between the ages derived from the cluster spectrum and the equivalent width of nebular H$\alpha$ strongly suggests that the H\two\ region is radiation-bounded.

\section{Other Clusters}
We fit population synthesis models to the STIS spectra of the other
four observed clusters using the method described above for M82-A1.
The quality of the spectra, however, is lower than for A1. In Table~\ref{param_clusters}, we present a summary of the derived parameters for the four clusters and compare them with previous work.

\subsection{Clusters F \& L}
The location and radial velocity of clusters M82-F and L (Gallagher
\& Smith 1999) places them within the 880 pc wide gaseous ring
surrounding the starburst core of the galaxy (Telesco et
al.\ 1991). The low velocity resolution of the STIS spectra does not allow us
to investigate the kinematics of the clusters. However, the relatively
high S/N and long wavelength coverage of the spectra allow us to use
synthesis techniques to ascertain the spectral parameters of the
clusters.

Figure~\ref{fig-F} shows the de-reddened, near-infrared spectrum of cluster F
compared to the best fit BC03 model.  (The SB99 models do not offer
good spectral resolution here.)  The quality of the fits is determined
by a weighted $\chi^{2}$ statistic.  We give more weight to spectral
regions containing the Ca\two\ triplet and the broad molecular TiO and VO
features.  We find that clusters F and L have ages of $50^{+30}_{-35}$
Myr and $65^{+70}_{-35}$ Myr respectively, with $E(B-V)$ values of
$0.64\pm0.10$ and $1.87\pm0.39$.  The largest source of uncertainty is
the fact that the spectral ageing of $\sim 20-100$ Myr old populations
in the $5000-10000\textrm{\AA}$ wavelength range can be mimicked by a
decrease in reddening.

Cluster F is associated with H$\alpha$ and [N\two] $\lambda6585$ line
emission.  Much of this is diffuse and kinematically unrelated to the
cluster, but we are unable to properly model this background because
it appears to be highly variable.  An upper limit to the emission line
equivalent width of H$\alpha$ in the spectrum of F is
$-0.3\pm0.2\textrm{\AA}$, from which the SB99 models predict a lower
limit to the cluster age of 22 Myr, consistent with the estimate
above.

Our age estimate for cluster F compares favourably with that of
Gallagher \& Smith (1999) who find a value of $60\pm20$ Myr (Table~\ref{param_clusters}).  
However, they derive a reddening value of
$\sim 0.9$, somewhat in excess of our estimate. This might be an
effect of patchy extinction in the vicinity of cluster F combined with
the lower spatial resolution of their ground-based spectrum. We do
notice that the slope of the cluster spectrum becomes progressively
redder towards the outskirts of the cluster, which might be taken as
evidence for a non-uniform dust screen or internal extinction
gradients within the cluster itself. This is consistent with the idea
that M82-F is seen through a hole in the projected dust distribution
of the galaxy (Gallagher and Smith 1999). As shown in Table~\ref{param_clusters}, McCrady et al. (2005)
find an even lower value for the reddening towards F based on ACS/HRC photometry.
\begin{figure*}
\psfig{file=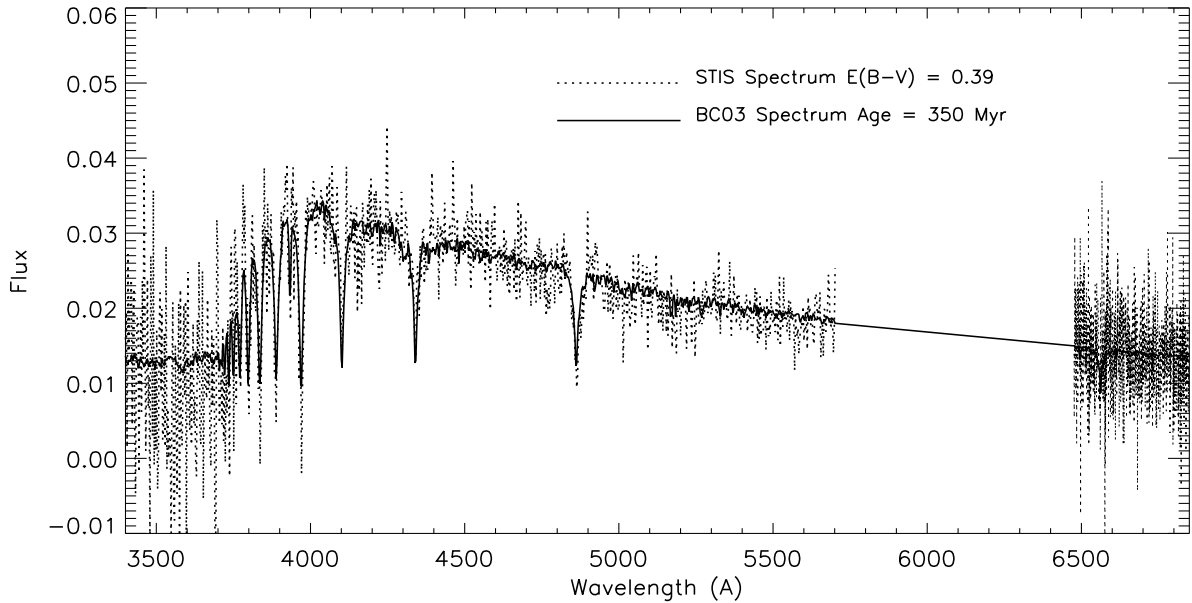,width=18cm,bbllx=20pt,bblly=20pt,bburx=370pt,bbury=750pt,angle=90}
\caption{Comparison of the de-reddened ($E(B-V)=0.39$) STIS spectrum of M82-B1-1 
with a synthetic BC03 spectrum for an age of 350 Myr.
The flux is in units of $10^{-14}$~erg~s$^{-1}$~cm$^{-2}$~\AA$^{-1}$}
\label{fig-B1}
\end{figure*} 
\begin{figure*}
\psfig{file=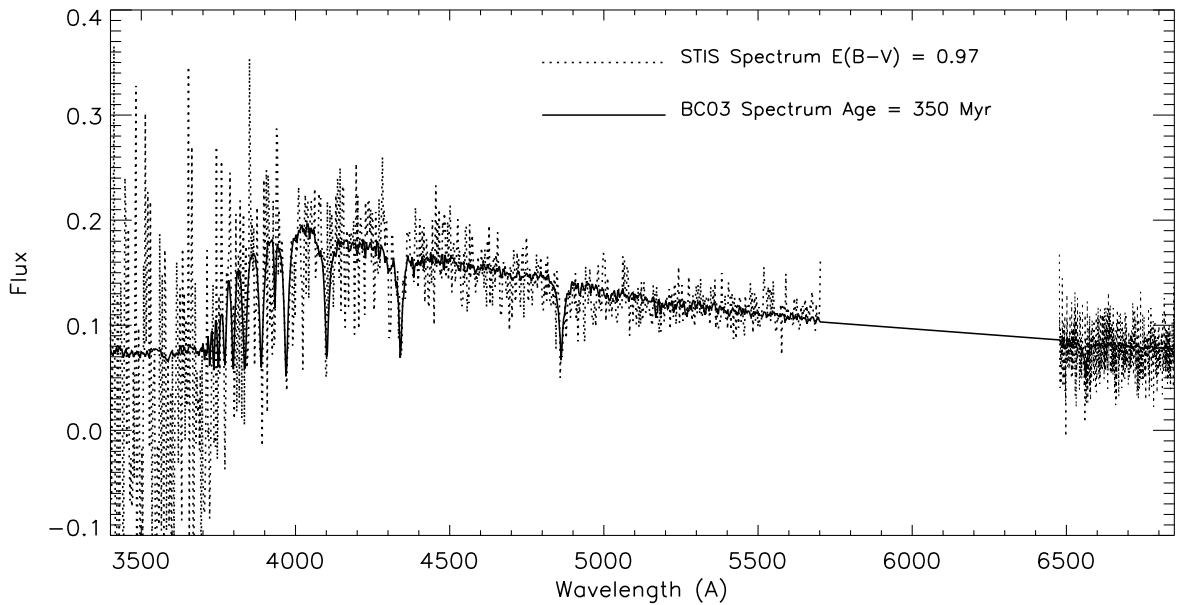,width=18cm,bbllx=20pt,bblly=20pt,bburx=370pt,bbury=750pt,angle=90}
\caption{Comparison of the de-reddened ($E(B-V)=0.97$) STIS spectrum of M82-B2-1
with a synthetic BC03 spectrum for an age of 350 Myr.
The flux is in units of $10^{-14}$~erg~s$^{-1}$~cm$^{-2}$~\AA$^{-1}$}
\label{fig-B2}
\end{figure*} 

Clusters F and L have comparable ages, which are distinct from those
in regions A and B.  Cluster L has considerably higher extinction
despite being near to F in projection.  Near infrared (1--2$\mu$m)
imaging does not reveal any other clusters in their vicinity (e.g.\
McCrady et al.\ 2003), so they do not seem to be members of a significant
intermediate age subsystem unless the foreground optical depth in dust
is very large. In addition, Gallagher \& Smith (1999) find that clusters F and L have quite different radial velocities of $+35$ and $+131$ \kms\ respectively.

Finally, we measured the spatial FWHM of the cluster F in different
spectral bins to look for any systematic decrease in the cluster size
towards redder wavelengths. After correcting for the known wavelength
dependent variation of the G750L PSF, we find no significant
secular changes in the size of the cluster.  We estimate an average
spatial FHWM of 3.6 pixels or about 183 mas. This compares favorably
to the average size of the cluster measured from ACS/HRC F814W images by
McCrady et al. (2005).
\begin{table}
\caption {Derived parameters for clusters M82-F, L, B1-1 and B2-1
and comparison with previous work}
\label{param_clusters}
\begin{tabular}{llll}
\hline
Cluster & Parameter & This Paper & Previous Work \\
\hline 
F & Age (Myr) & $50^{+30}_{-35}$ & $60\pm20^1$\\
   &			   &                        & 40--60$^2$\\
&    $E(B-V)$ & $0.64\pm0.10$ & $0.9\pm0.1^3$\\
& & & 0.19--0.37$^2$\\
L & Age (Myr) & $65^{+70}_{-35}$ & $\sim 60^1$\\
&    $E(B-V)$ & $1.87\pm0.39$ & $\ge 2^3$\\
B1-1 & Log (age) (yr)& $8.54\pm0.4$ & 8.9$^4$\\
& & & 8.86$^5$\\
&  $E(B-V)$ & $0.39\pm0.16$ & $0.13^4$\\
& & & 0.04$^5$\\
B2-1 & Log (age) (yr)& $8.54\pm0.8$ & 9.7$^4$\\
& & & 9.76$^5$\\
&  $E(B-V)$ & $0.97\pm0.51$ & $0.22^4$\\
& & & 0.00$^5$\\
\hline
\end{tabular}

$^{1}$Gallagher \& Smith (1999); $^2$McCrady et al. (2005); $^3$Smith \& Gallagher (2001); $^4$de Grijs et al. (2001); $^5$de Grijs et al. (2003).
\end{table}

\subsection{Clusters B1-1 and B2-1}
In Figures~\ref{fig-B1} and \ref{fig-B2}, we present the spectra and best-fit models for
the two brightest clusters in regions B1 and B2. Both clusters show strong
Balmer absorption lines and a deep Balmer jump, which are signatures
of an intermediate age star cluster.  The individual cluster spectra
are typical of the integrated spectrum of region B and most of the
rest of the main body of M82 outside the starburst core (described in
de Grijs et al.\ 2001).  (Seen in a distant galaxy, this would be
called a post-starburst or `E+A' type spectrum.)  From our fitting
routines, we derive log[age(yr)]$=8.54\pm0.4$ for B1-1 and
log[age(yr)]$=8.54\pm0.8$ for B2-1.  The significantly higher
uncertainty for the second cluster is a combination of the larger
extinction (E(B-V)$=0.97\pm0.51$, as opposed to E(B-V)$=0.39\pm0.16$
for B1-1) as well as a lower average S/N.

These results are generally consistent with those from ground-based
spectrophotometry and HST imaging of region B, which contains a large
number of intermediate age SSCs (Marcum \& O'Connell 1996, de Grijs et
al.\ 2001), probably associated with an intense star formation burst
during the last close passage of M81 and M82 about 1 Gyr ago.
However, as shown in Table~\ref{param_clusters}, our ages are younger and extinctions higher than those for
these clusters based on WFPC2 $BVI$ photometry by de Grijs et al.\
(2001, 2003).  We believe
our spectroscopic ages are more robust, owing to the difficulty of
doing aperture photometry in such a complex region.

\section{Discussion}
The combination of high spatial resolution HST/STIS optical spectra and
improved spectral synthesis model spectra have allowed us to derive
numerous parameters for the young massive cluster M82-A1 and its
H\two\ region in the starburst core of M82. These parameters are
summarised in Table~\ref{param}. 
\subsection{M82-A1 and the Evolution of the M82 Starburst}
We now discuss the age and location of cluster M82-A1 in terms of the
history of the central starburst.  As discussed in Section 1,
O'Connell et al. (1995) put forward various arguments to suggest that
region A (with C and E) is the part of the starburst core that is
least obscured along the line of sight. Cluster M82-A1 is located
130~pc north-east of the $2\mu m$ nucleus, slightly beyond the eastern
edge of the 100~pc radius ionized arc identified by Achtermann \&
Lacey (1995) from [Ne\two] observations. Its measured radial velocity
of $+320\pm20$ \kms\ agrees well with the position-velocity maps of
Achtermann \& Lacey (1995).  The cluster thus appears to be spatially
and kinematically associated with the ionized gas seen in the
starburst core of M82.

The evolutionary synthesis models of F\"orster Schreiber et al. (2003)
reproduce the observed characteristics of the M82 starburst if there
were two short duration bursts occurring 8--15 and 4--6 Myr ago. The
first burst was concentrated in the central few tens of parsecs and
the second episode occurred mainly in the circumnuclear regions. On
the basis of its position and age of 6.4 Myr, we associate M82-A1 with
the more recent burst. Melo et al. (2005) catalogue 86 clusters in
region A and find this region has the highest projected surface
density of young clusters. The morphology of region A
(cf. Fig.~\ref{fig-slit}) suggests the clusters are coeval. This is
strengthened by the fact that the bright base (in projection) of the
superwind is centred on regions A and C (O'Connell \& Mangano 1978;
Shopbell \& Bland-Hawthorn 1998; Gallagher et al., in prep.),
indicating that the clusters must be of a similar age since the cluster winds 
are very time dependent. If this is the
case, then region A must correspond to a particularly intense episode
of star formation $\approx 6$~Myr ago.

\subsection{The M82-A1 H\two\ Region}
We now discuss the compact H\two\ region associated with M82-A1.
F\"orster Schreiber et al. (2001) have studied the physical conditions of
the ISM in the starburst core of M82 from ground and space-based
IR observations. They find that a random distribution of closely-packed
ionizing clusters and small gas clouds, separated on average by a few parsecs,
represents well the starburst regions of M82. They derive a
constant ionization parameter $\log U$ of $-2.3$  on spatial scales of a few tens of parsecs to 500~pc, and suggest that this uniform degree of ionization indicates a similar star formation efficiency and evolutionary stage for the M82 starburst regions.

The H\two\ region around M82-A1 represents a photoionized nebula
on a much smaller physical scale than previous observations of 
F\"orster Schreiber et al. (2001) which probe
down to structures typically on 25~pc scales. For comparison with
their model, we derive the
ionization parameter for the M82-A1 H\two\ region as given by:
\begin{equation}
 U = \frac {Q_{0}} {4 \pi R^{2} N_{\rm e}\ c} 
\end{equation}
where $Q_{0}$ represents the photon luminosity in the Lyman
continuum, $R$ is the nebula radius, and \Ne\ is the electron density. Taking
the values given in Table~\ref{param}, we obtain $\log U =-2.24\pm0.18$,
in excellent agreement with the value of $-2.3$ derived by
F\"orster Schreiber et al. (2001), and thus extending the uniformity of the
M82 starburst down to scales of a few parsecs or individual cluster H\two\ regions.
This suggests that the whole M82 starburst region is self-adjusting in the sense 
that the number of ionizing photons per atom is constant. 

The value of $U$ derived by
F\"orster Schreiber et al. (2001) assumed an electron density of 300~\cm3, which
is far lower than the value we find for the M82-A1 H\two\ region and the
surrounding medium. We suspect these differences are due to sampling
different environments; the IR observations of F\"orster Schreiber et al. (2001)
probably probe deeper into the starburst region, and into a region where the
gas is less compressed.

Turning now to the size of the H\two\ region,
we find that it is compact with a radius only 1.5 times the cluster radius.
It has a high pressure of $P/k = 1$--$2 \times 10^{7}$~\cm3\,K.
The ambient thermal pressure of the diffuse ionized gas in region A is also high with
$P/k \sim 10^{7}$~\cm3\,K. The total ambient pressure is probably even greater than this. The STIS spectra of the diffuse gas in region A show that the H$\alpha$ emission is broad with an average velocity dispersion $\sigma_{\rm v}$ of 45 km\,s$^{-1}$, giving $P_{{\rm turb}}/k \simeq \rho\, \sigma_{\rm v}^{2}/k \sim 3 \times 10^{8}$~\cm3\,K.
This suggests that the turbulent pressure component dominates. It presumably originates from the nonthermal energy input from supernovae events.
The M82-A1 H\two\ region is therefore in an unusually high pressure environment which will control its evolution.

Observations of radio supernova remnants (Muxlow et al. 1994; Pedlar et al. 1999) in M82 show that they are compact (with radii $< 4$ pc) for their inferred ages; these authors suggest that their growth has been inhibited by the dense starburst environment in M82. Chevalier \& Fransson (2001) propose that these sources are evolved supernova remnants interacting with a dense interstellar medium; they may have recently entered the radiative phase if the surrounding density is $\approx 1000$ \cm3. We find that the density in region A is of this order, and that the compact H\two\ region surrounding M82-A1 has a similar size to the upper limit on the sizes of the radio supernova remnants. We thus postulate that the compactness of both the SNRs and the M82-A1 H\two\ region is the result of the high ambient pressure in the starburst core.

The continuous injection of energy by winds into the interstellar medium results in the formation of an adiabatic, pressure-driven bubble (e.g. Weaver et al. 1977; Chevalier \& Clegg 1985). For a star cluster, the source of energy is first stellar winds and then supernovae. For M82-A1, we estimate that its cluster wind has a mechanical luminosity $L_{\rm{mech}} \approx 2.5 \times 10^{40}$ erg\,s$^{-1}$ from the Starburst99 models for a Kroupa IMF and the derived mass of M82-A1.
We can estimate the radius of the shell formed by the cluster wind using the standard equation for pressure-driven superbubbles from Weaver et al. (1977):
\begin{equation}
R = \left( \frac{125}{154\pi}\right)^{1/5}\ 
\left(\frac{L_{\rm{mech}}}{\rho_{0}}\right)^{1/5} t^{3/5}
\end{equation}   
where $R$ is the shell radius, $\rho_{0}$ is the ambient mass density, and $t$ is the age. Using the value of
$L_{\rm{mech}}$ given above, an ambient
number density of 1000~\cm3 (Sect.~\ref{para-hii}), and an age of 6.4~Myr, the predicted
radius is 150 or 100 pc, if we assume a thermalisation efficiency of 100\% or 10\% respectively. These values are obviously far in excess of the measured radius of 4.5~pc, and suggest that the H\two\ region has not developed according to the standard model.

Oey \& Garc\'\i a-Segura (2004) and Dopita et al. (2005) consider the effect of the ambient interstellar pressure on the dynamical evolution of pressure-driven superbubbles. A high ambient pressure will inhibit the growth of the H\two\ region and it will stall when the internal pressure of the superbubble equals the ambient pressure. The internal pressure $P$ is given by Weaver et al. (1977) as:
\begin{equation}
P = \frac{7}{(3850\pi)^{2/5}}\left( \frac{125}{154\pi}\right)^{4/15}\ \left(\frac{L_{\rm{mech}}}{\rho_{0}}\right)^{2/3}\ \frac{\rho_{0}}{R^{4/3}}.
\end{equation}
This leads to an internal pressure $P/k$ of $\sim 3 \times 10^{9}$ or $\sim 6 \times 10^{8}$~\cm3\,K, assuming thermal efficiencies of 100 and 10\% respectively. The latter value is reasonably close to our estimate of the total (thermal and turbulent) ambient interstellar pressure of $\sim 3\times 10^{8}$~\cm3\,K. While this is a rough estimate, and does not take into account the fact that the bubble will not abruptly stop expanding when it stalls, it illustrates that the high ambient pressure is probably sufficient to modify the evolution of the H\two\ region surrounding the cluster M82-A1. 

Using Eqn. (3), Dopita et al. (2005) show that all stalled H\two\ regions will have a common ratio of
$L_{\rm{mech}}/\rho\,R^{2}$, and since $L_{\rm{mech}}$ scales roughly with the flux of ionizing photons $Q_{0}$, then all stalled H\two\ regions should have the same ionization parameter as given by Eqn. (1).
This relationship may help to explain the uniformity of the ionization parameter on many scales in the starburst regions of M82 discussed above and by F\"orster Schreiber et al. (2001).

Another important possibility to consider is whether the cluster wind from M82-A1 has
developed according to the standard adiabatically expanding wind model. Canto, Raga \& Rodr\'\i guez (2000) present an analytic mass-loaded cluster wind model for a dense cluster of massive stars. They find that there are two solutions depending on the properties of the surrounding ISM; the standard supersonic solution or a subsonic solution if the environment is at high pressure. The latter solution may apply to M82-A1. 
The supersonic wind solution can also be inhibited if the cluster wind is sufficiently dense to promote radiative cooling (Silich, Tenorio-Tagle \& Mu\~ non-Tu\~n\'on 2003). This occurs if the cluster is massive but compact and a supernebula can then be formed
(Tenorio-Tagle et al. 2005). From their figures, we find that the cluster M82-A1 is not in the regime where catastrophic cooling will occur.

 Ultimately, it is hard to understand how the observed M82 galactic superwind is generated if the environment stifles its development because of the high ambient pressure. Chevalier \& Fransson (2001) consider this problem in relation to their prediction that the compact radio supernovae in M82 are radiative and thus are unable to provide sufficient energy to drive a galactic wind. They hypothesise that galactic winds are driven only by supernovae occurring in a low density, hot medium. Another way around this problem is to conjecture 
that there is a pressure gradient in the central region of M82 such that the outer clusters are in a lower pressure region. The winds from these clusters will then be able to develop and may produce highly-collimated outflows, as envisaged in the interacting wind model of Tenorio-Tagle, Silich \& 
Mu\~ non-Tu\~n\'on (2003).

Whatever the precise reason for the compact nature of the H\two\ region surrounding M82-A1, it is clear that the unusually high ambient pressure in the M82 starburst environment must play an important role. 
\section{Summary and Conclusions}

We have presented optical HST/STIS spectroscopy of five massive
clusters in M82.  Our best data are for cluster M82-A1 and its
environment in region A of the starburst core. This region
contains a large complex of young clusters which appear to be
responsible for a major component of the superwind. We have sought to
derive the properties of M82-A1, and thereby clarify the age of the
recent starburst, and to investigate the nature of its ionized
environment.

The STIS G430L spectrum of M82-A1 is dominated by Balmer absorption
lines and a weak Balmer jump suggesting a young age. We measure a mean
heliocentric velocity of $+320\pm20$ km\,s$^{-1}$. This value and its
position 130 pc north-east of the 2$\mu$m nucleus suggests that M82-A1
is spatially and kinematically associated with the starburst core.

Using the evolutionary synthesis
code Starburst99 (Leitherer et al. 1999) which includes a high
resolution spectral library to derive both the age and reddening of
M82-A1, we find that we can only match the spectral energy
distribution and depth of the Balmer jump for a narrow age range of
5--7 Myr and a reddening of $E(B-V)=1.5$--1.2 mag.  The independent
BC03 models give similar results. 

The STIS G750M spectrum shows that M82-A1 is surrounded by a compact
H\two\ region.  We have recovered the intrinsic nebular spectrum by
subtracting the best-fitting synthetic spectrum and find that the
nebular reddening of $E(B-V)=1.38\pm0.15$ mag agrees well with the
value determined from the cluster SED. We adopt a mean of the two
methods to give $E(B-V)=1.35\pm0.15$ mag. The observed equivalent
width of the H$\alpha$ emission line provides an upper limit to the
age of 6.7 Myr, in agreement with the depth of the Balmer jump. We
adopt an age of $6.4\pm0.5$ Myr for M82-A1.

Comparison with the models of F\"orster Schreiber et al. (2003)
suggest that the formation of M82-A1 is associated with the most
recent burst of star formation which took place $\approx$4--6 Myr
ago. The overall morphology of region A and the fact that a major
component of the superwind is centred on this region suggests that the
clusters are coeval. If this is the case, then region A must
correspond to a particularly intense episode of star formation.

We have used archive HST images to derive the radius and mass of
M82-A1 via photometry. We find that the cluster is elliptical with a
mean effective radius of $3.0\pm0.5$ pc and a minor/major axis ratio
of $0.72\pm0.05$. The cluster effective radius appears to decrease
towards shorter wavelengths. We interpret this behaviour as being due to dust in the surrounding H\two\ region rather than mass segregation because it seems difficult to produce a radial colour gradient in a young cluster.
We derive a mass of 0.7--1.3$\times 10^{6}$
M$_{\odot}$ for M82-A1 depending on whether we adopt a Kroupa or
Salpeter IMF.

We have sought to understand the nature of the H\two\ region
surrounding M82-A1. From HST/PC images, we have derived an effective
radius of only $4.5\pm0.5$ pc and find that it is highly elliptical
with a minor/major axis ratio of 0.57. It is also of high density; the
[S\two] doublet in the STIS G750M spectrum gives an electron density
of $1800\pm300$ cm$^{-3}$ or a thermal pressure of $P/k = 1$--$2 \times
10^{7}$~\cm3\,K, for \Te=$0.5$--1$\times 10^4$.  
The diffuse ionized gas along the STIS slit has a
slightly lower density which, together with a measured high turbulent
velocity dispersion of 45 km\,s$^{-1}$, suggests that M82-A1 is
immersed in an unusually high pressure environment. We also find that
the oxygen abundance of the H\two\ region is close to, or slightly
higher than, solar.

We find that the ionization parameter $\log U$ of the H\two\ region is
$-2.24\pm0.18$ in excellent agreement with the value of $-2.3$ derived
by F\"orster Schreiber et al. (2001) for scales of a few tens to 500
pc. Our measurement thus extends the uniformity of the M82 starburst
down to the scale of individual cluster H\two\ regions, and suggests
that the entire starburst is self-adjusting with a constant number of
ionizing photons per atom.

We have sought to understand the compactness of the H\two\
region around M82-A1, given that it is much smaller than the size
predicted for its age according to the standard model for a pressure-driven
superbubble. We find that the high pressure of the environment may
have caused the bubble to stall or follow a different wind solution to
the standard supersonic adiabatic one. We suggest that models should
be developed which account for the evolution of a cluster wind in a
high pressure environment.

We also derive spectroscopic ages and reddening for four other
clusters in M82 (B1-1, B2-1, F and L).  These are consistent with
earlier studies and demonstrate that star formation activity,
sufficiently intense to produce super star clusters, has been going on
in M82 during the past Gyr, perhaps in discrete and localized
episodes.

\section*{Acknowledgments}

We thank David Stys and Linda Dressel at STScI for their help in
reducing the STIS observations, and Mario Livio for useful discussions.
LJS and MSW thank the Department of
Astronomy at the University of Wisconsin-Madison for their warm
hospitality, and the Wisconsin and UCL Graduate Schools for financial
support during the writing of this paper. We also thank Marcello Mora
for the ACS PSF data and Nate Bastian for his help in using {\sc
ishape}.  This paper is based on observations with the NASA/ESA {\it
Hubble Space Telescope}\/ which is operated by the Association of
Universities for Research in Astronomy, Inc. under NASA contract
NAS5-26555.  It made use of HST data taken as part of GO programs
6826, 7218, 7446, 9117 and 9788.  This research was supported in part by
NASA LTSA grant NAG5-6403 and STScI grant GO-09117 to the University
of Virginia. The Image Reduction and Analysis Facility ({\sc iraf}) is
distributed by the National Optical Astronomy Observatories which is
operated by the Association of Universities for Research in Astronomy,
Inc.  under cooperative agreement with the National Science
Foundation.  {\sc stsdas} is the Space Telescope Science Data Analysis
System; its tasks are complementary to those in {\sc iraf}.

\bsp
\label{lastpage}
\end{document}